\documentclass[10pt]{article}
\usepackage{amsmath,amsfonts,latexsym,amsthm,amssymb}
\numberwithin{equation}{section}
\newtheorem{theorem}{Theorem}[section]
\newtheorem{proposition}{Proposition}[section]
\newtheorem{lemma}{Lemma}[section]
\newtheorem{definition}{Definition}[section]

\newcommand\DCRn{{\mathcal D}'({\mathbb R}^n)}
\newcommand{\DCQpn}{{\mathcal D}'({\mathbb Q}_p^n)}
\newcommand\DRn{{\mathcal D}({\mathbb R}^n)}
\newcommand\DO{{\mathcal D}({O})}
\newcommand\DCO{{\mathcal D}'({O})}
\newcommand\DK{{\mathcal D}(K)}
\newcommand\DlPi{{\mathcal D}_l(\Pi)}

\newcommand\DCK{{\mathcal D}'(K)}
\newcommand\DPi{{\mathcal D}(\Pi)}
\newcommand\DCPi{{\mathcal D}'(\Pi)}
\newcommand\DClPi{{{\mathcal D}_l}'(\Pi)}
\newcommand\SCRn{{\mathcal S}'({\mathbb R}^n)}

\newcommand\DQpn{{\mathcal D}({\mathbb Q}_p^n)}
\newcommand\Rn{{\mathbb R}^n}
\newcommand\Qpn{{\mathbb Q}_p^n}
\newcommand\Qp{{\mathbb Q}_p}
\newcommand\lug{\langle}
\newcommand\rug{\rangle}
\newcommand\vphi{\varphi}
\newcommand\vphid{\varphi_\delta}
\newcommand\vk{\varkappa}
\newcommand\al{\alpha}
\newcommand\supp{\mathop{\rm supp}}
\newcommand{\wick}[1]{\colon\hskip-2.2pt#1\hskip-0.5pt\colon\hskip-2pt}
\newcommand\EE{{\mathcal E}}
\renewcommand\AA{{\mathfrak A}}
\newcommand\HH{{\mathcal H}}
\newcommand\APi{{\mathfrak A}_{\Pi}}
\newcommand\AtPi{\widetilde{\mathfrak A}_{\Pi}}
\newcommand\wh{\widehat}
\newcommand\SchdgPi{S_{\delta,\,g}^{(\Pi)}}
\newcommand\SchdgPiC{S_{\delta,\,g}^{(\Pi')}}
\newcommand\dmuPi{d\mu_0^{\Pi}}
\begin{document}

\centerline{\bf INTERACTION MEASURES ON THE SPACE OF
DISTRIBUTIONS}

\centerline{\bf OVER THE FIELD OF p-ADIC NUMBERS}

\bigskip

\centerline{\footnotesize ANATOLY N. KOCHUBEI and MUSTAFA R.
SAIT-AMETOV}

\baselineskip=12pt \centerline{\footnotesize\it Institute of
Mathematics}

\baselineskip=12pt \centerline{\footnotesize\it National Academy
of Sciences of Ukraine}

\baselineskip=12pt \centerline{\footnotesize\it Tereshchenkivska 3, Kiev 01601,
Ukraine}

\baselineskip=10pt \centerline{\footnotesize\it e-mail:
kochubei@i.com.ua, mustafa@imath.kiev.ua}

\vspace*{1truein}

{{\bf Abstract.}\ We construct measures on the space $\DCQpn$, $n\le 4$, of
Bruhat-Schwartz distributions over the field of $p$-adic numbers, corresponding to finite volume
polynomial interactions in a $p$-adic analog of the Euclidean quantum field theory. In contrast
to earlier results in this direction, our choice of the free measure is the Gaussian measure
corresponding to an elliptic pseudo-differential operator over $\Qpn$. Analogs of the Euclidean
$P(\vphi)$-theories with free and half-Dirichlet boundary conditions are considered.}{}{}

\vspace*{4pt}

\section{Introduction}    
\noindent \looseness1 The basic objects of the Euclidean quantum
field theory~\cite{GJ,M,Si} are probability measures on the space
$\DCRn$ or $\SCRn$ of real distributions. Equivalently, one can
speak about a generalized random process $\vphi(f)$, $f\in\DRn$,
fixing a probability measure $\mu_0$ on the Borel $\sigma$-algebra
$\Sigma$ of $\DCRn$ and considering the probability space
$(\DCRn,\Sigma,\mu_0)$. The free boson field is described by the
measure $\mu_0$ corresponding to the Gaussian process with mean
zero and covariance $\lug\,\vphi (f)\,\vphi
(g)\,\rug=\left(f,(-\Delta +m^2)^{-1}g\right)$. In order to
describe fields with interaction corresponding to a polynomial
$P$, it is necessary to define $P(\vphi)$ via some renormalization
procedure. Then, for any $g\in\DRn$, the measure
\begin{equation}
\label{eqn1.1} d\mu_g (\vphi) = \frac{\exp \left\{ -\left( P(\vphi),g \right) \right\} d\mu_0
(\vphi) }{ \int\exp \left\{ -\left( P(\vphi),g \right)\right\} d\mu_0 (\vphi)}\ ,
\end{equation}
if the expression (1.1) makes sense and indeed defines a measure,
is interpreted as an interaction measure in a finite volume. In
some cases there exists (in a certain sense) also an infinite
volume limit $\lim_{g\to 1} d\mu_g$.

Within the recent tendency to find non-Archimedean analogs of all
important objects of mathematical physics,\cite{Kh,K,VVZ} it is
natural to look for $p$-adic counterparts of the above
constructions. This problem (formulated in Refs. 14, 16) is of a
clear mathematical interest as a major problem of the
infinite-dimensional non-Archimedean analysis, irrespective of
possible physical applications.

As we switch from $\Rn$ to $\Qpn=\Qp\times\cdots\times\Qp$, where $\Qp$ is the field of $p$-adic
numbers, we consider the space $\DCQpn$ of Bruhat-Schwartz distributions. It turns out that the
first crucial step is the choice of a free covariance. Note that one cannot define differential
operators acting on complex-valued functions over $\Qp$.

The first results in $p$-adic quantum field theory were obtained
in a series of papers by Lerner and Missarov (see
e.g. Refs. 7, 8, 10) whose main motivation was to find a
continuous analog of hierarchical models. Their choice of  a free
covariance was $$\text{\rm const}\int_{\Qpn\times\Qpn} f(x)
\|x-y\|^{-\al n} f(y)\,dx\, dy,\qquad \al>0,$$ which had nice
scaling and discretization properties, made it possible to define
a Hamiltonian of the discretized interacting field, but did not
lead to the construction (\ref{eqn1.1}).

In this paper we propose a different approach. We follow the
construction of Refs. 13, 2 using (instead of the Laplacian)
elliptic pseudo-differential operators over $\Qp$ introduced
in Refs 6,5. Such elliptic operators exist only for $n\le4$,
and in the ``physical'' dimension 4 such an operator is unique up
to an isomorphism. Below we assume that $p\ne2$.

More specifically, let $h(\xi_1,\ldots,\xi_n)$ be a quadratic form
with coefficients from $\Qp$, such that the ellipticity  (or
anisotropy) condition
\begin{equation}
\label{eqn1.2}
h(\xi_1,\ldots,\xi_n)\ne0\qquad{\rm
if}\qquad |\xi_1|_p+\cdots+|\xi_n|_p\ne0
\end{equation}
holds. On the space $L_2(\Qpn)$ of square integrable
complex-valued functions with respect to the additive Haar measure
we consider the self-adjoint positive operator
$A=F^{-1}M_{h,\al}F$, where $M_{h,\al}$ is the operator of
multiplication by $|h(\xi_1,\ldots,\xi_n)|_p^{\al}$, $\al>0$, $F$
is the Fourier transform (for the main notions and results of
non-Archimedean analysis used in this paper see Ref. 5)

If $O$ is a ball in $\Qpn$ (with respect to the absolute value
$\max_{1\le j\le n} |\xi_j|_p$), then $A_{O}$ will denote the
operator on $L_2(O)$ defined as follows. Let $f\in\DO$, that is
$f\in\DQpn$, $\supp{f} \subset O$. Extending $f$ onto $\Qpn$ by
zero, we apply the operator $A$ to that extension. Restricting the
resulting function to $O$ we obtain a function from $L_2(O)$ which
is taken as $A_Of$. This defines an operator on $\DO$; its closure
$A_O$ is a self-adjoint positive operator on $L_2(O)$.

Let $\mu_0$ be the measure on $\DCQpn$ corresponding to the
Gaussian process with the mean zero and covariance
$\lug\,\vphi(f)\,\vphi(g)\,\rug=\left(f,(A+m^2)^{-1}g\right)$,
where $(\cdot,\cdot)$ is the inner product in $L_2(\Qpn)$. As it
is dictated by the condition (\ref{eqn1.2}), we always assume
that $n\le4$. We show that for any semibounded polynomial $P$ and
any $\al\ge\frac{n}{2}$, we can define $P(\vphi)$ to be the Wick
renormalization with respect to the measure $\mu_0$, as it is done
in the Euclidean $P(\vphi)_2$-model.\cite{Si,GJ} The resulting
non-Gaussian generalized stochastic process will be denoted
$\wick{P(\vphi)}$. Moreover, with this renormalization the measure
(\ref{eqn1.1}) is well-defined. It may be seen as a $p$-adic
counterpart of the Euclidean $P(\vphi)_2$ with free boundary
conditions.\cite{Si}

Another option (resembling the $P(\vphi)_2$ with the
half-Dirichlet boundary conditions from Ref. 13) is to use the
above Wick renormalization with respect to $\mu_0$ while replacing
$\mu_0$ in (\ref{eqn1.1}) by $\mu_0^O$, the measure on $\DCO$
corresponding, as above, to the operator $A_O$, and taking $g$ to
be the indicator function of the ball $O$. We show that
(\ref{eqn1.1}) makes sense in this case too. This approach has
some preferences. Though the infinite volume limit is not
considered in this paper, we prove that within this ``mixed''
construction of the interaction measure the corresponding
Schwinger functionals (under some assumptions on $P$) depend
monotonously on the radius of the ball $O$. This result is based
on a version of the lattice (or, rather, graph) approximation,
which is of some independent interest.

\medskip
\section{Elliptic operators}
\noindent \looseness1 In this section we introduce the elliptic
pseudo-differential operators that play the role of the Laplacian
in our approach. Hereinafter we will assume that $m$ is some fixed
positive constant, $\alpha \ge \dfrac{n}2$.

\subsection{Basic information}
\noindent The theory of the operator $A$ defined in the
introduction is expounded in detail in Ref. 5. The results
needed here can be summarized as follows.

The resolvent $(A+m^2)^{-1}$ is an integral operator of
convolution type, $$ \left( (A+m^2)^{-1} f\right) (x)=\int_{\Qpn}
\EE (X-Y)f(Y)dY,\qquad X\in\Qpn. $$ The investigation of the Green
function $\EE(X)=\EE(x_1,\ldots,x_n)$, $x_1,\ldots,x_n\in\Qp$, is
based on a procedure of reduction of multi-dimensional
pseudo-differential operators over $\Qp$ to one-dimensional
operators on more general fields.

The vector space $K=\Qpn$ ($n\le4$) can be endowed with additional
algebraic structures, so that $K$ is assumed to be a local field
(an extension of $\Qp$), if $n\le3$, or the non-commutative
quaternion algebra over $\Qp$, if $n=4$. Let $\|\cdot\|$ be the
normalized absolute value, $\beta$ be a prime element of $K$, $q$
be the cardinality of the residue field of $K$. For $x\in K$,
$\|x\|=q^N$, we will write $x=\beta^{-N}u_{x}$,   $\|u_x\|=1$.

It is shown in Ref. 5 that for each quadratic form $h$
satisfying (\ref{eqn1.2}) we can construct $K$ in such way
that
\begin{equation}
\label{eqn2.1}
\EE(x_1,\ldots,x_n)=\vert {\rm det}\ T\vert_p^{-1} E\left(
({T'}^{-1}X)_1e_1+\cdots+({T'}^{-1}X)_ne_n)\right)
\end{equation}
where $e_1,\ldots,e_n$ is a basis of $K$ over $\Qp$, $T\in
GL(n,\Qp)$, $T'$ is the transpose of $T$, $E$ is the Green function, that
is the integral (convolution) kernel of the resolvent $(\AA +
m^2)^{-1}$ of a pseudo-differential operator $\AA$ over $K$ with
the symbol
\begin{equation}
\label{eqn2.2}
a(\xi)=\|\xi\|^{2\al/n}\gamma(u_{\xi}),\qquad \xi\in K,
\end{equation}
where $\gamma$ is a continuous strictly positive function on the
group of units $U$ of $K$. As usual, the operator $\AA$ is defined
on the space $\DK$ of locally constant functions with compact
supports as $\AA=F_K^{-1}M_aF_K$, where $M_a$ is the operator of
multiplication by $a$, $F_K$ is the Fourier transform, that is $$
(F_Kf)(\xi)=\int_K \chi(\xi x)f(x)\,dx,\qquad \xi\in K, $$ where
$\chi$  is a rank zero additive character on $K$. Below we will
often use the notation $\widehat{f}=F_Kf$.

Explicit forms of all the above objects (the choice of the
expressions for function $\gamma$, the basis of coordinate
representation $e_1,\ldots,e_n$, the matrix $T$ appearing in
(\ref{eqn2.1}) etc.) can be found for various classes of the
forms $h$ in Ref. 5.

The relation (\ref{eqn2.1}) shows that the Gaussian measure
$\mu_0$ corresponding to the operator $A$ and a similar measure
constructed from the operator $\AA$ (we can identify $K$ and
$\Qpn$ via the basis $\{e_j\}$) are connected by a simple linear
transformation. Therefore we will substantiate (\ref{eqn1.1})
using $\AA$ instead of $A$. In order to simplify the notations,
below we will ignore the difference between $A$ and $\AA$, and
understand $\mu_0$ as the measure corresponding to $\AA$.

Under the above identification of $K$ and $\Qpn$ a ball $O$ in
$\Qpn$ corresponds to a compact set $\Pi$ in $K$ which is not
necessary a ball, but the union of a finite number of disjoint
balls. Therefore, in the construction of the mixed interaction
measure instead of $\mu_0^O$ (defined in the introduction) we will
deal with a similar measure $\mu_0^\Pi$.

Later we will often make use of the following technical lemma.
\smallskip\nobreak
\begin{lemma}
Let the function $a(\xi)$ be as in (\ref{eqn2.2}). If $\al\ge
\frac{n}{2}$, then for any $\vk\in {\mathbb Z}$, $\vk \ge 1$,
\begin{equation}
\label{eqn2.3} \int_{\|\xi\|\le q^\vk}
\left(a(\xi)+m^2\right)^{-1}\,d\xi\ \le\ c_1\vk,
\end{equation}
and for any $\beta>1$
\begin{equation}
\label{eqn2.4} \int_{\|\xi\|\ge q^\vk}
\left(a(\xi)+m^2\right)^{-\beta}\,d\xi\ \le\
c_2\,q^{-\vk(2\al\beta/n-1)},
\end{equation}
where $c_1$ and $c_2$ are positive constants that do not depend on
$\vk$.
\end{lemma}
\begin{proof}
First, note that $\gamma(u_\xi)$ in the definition
(\ref{eqn2.2}) is a positive continuous function on a compact
set. Thus there exist such positive constants $\gamma_{\min}$ and
$\gamma_{\max}$, that $$\gamma_{\min}\ \le\ \gamma(u_\xi)\ \le\
\gamma_{\max}\quad \text{for any}\quad \xi\in K.$$ The following
simple calculations prove (\ref{eqn2.3}):
\begin{multline*}
\int_{\|\xi\|\le q^\vk} \left(a(\xi)+m^2\right)^{-1}\,d\xi \le
m^{-2}\int_{\|\xi\|\le 1} \,d\xi\ +\
\gamma_{\min}^{-1}\int_{1<\|\xi\|\le q^\vk}
\|\xi\|^{-2\al/n}\,d\xi\ \\ \le m^{-2}\ +\
\gamma_{\min}^{-1}(1-q^{-1})\sum_{l=1}^{\vk} q^{-l(2\al/n\ -1)}\
\le\ m^{-2} + \gamma_{\min}^{-1}(1-q^{-1})\,\vk.
\end{multline*}
Next,
\begin{multline*}
\int_{\|\xi\|\ge q^\vk} \left(a(\xi)+m^2\right)^{-\beta}\,d\xi\
\le\ \gamma_{\min}^{-1} \int_{\|\xi\|\ge q^\vk}
\|\xi\|^{-2\al\beta/n}\,d\xi\ \\ 
\le \gamma_{\min}^{-1}(1-q^{-1})\sum_{l=\vk}^{\infty}
q^{-l(2\al\beta/n\ -1)}\ \le\ c_2\,q^{-\vk(2\al\beta/n\ -1)},
\end{multline*}
since $\frac{2\al\beta}{n}>0$ and $q>1$.
\end{proof}

Calculations similar to those in proof of (\ref{eqn2.3}) give
the estimate
\begin{equation}
\label{eqn2.5} \int_{\|\xi\|\le q^\vk}
\left(a(\xi)+m^2\right)^{-\beta}\,d\xi\ \le\ c\,\vk,
\end{equation}
where positive constant $c$ does not depend on $\vk$. Thus from
(\ref{eqn2.4}) and (\ref{eqn2.5}) we can conclude that the
integral $\int_K \left(a(\xi)+m^2\right)^{-\beta}\,d\xi$ converges
for any $\beta>1$.

\subsection{Properties of the operator $\AA$ and its resolvent}
\noindent \looseness1 Let us list some properties of the operator
$\AA$ and its Green function E. For the proofs see Ref. 5.

The operator $\AA$ admits a hyper-singular integral representation
\begin{equation}
\label{eqn2.6} \left(\AA z\right)(x)=\int_K \|y\|^{-2\al/n -1}
\Omega(u_y)[z(x-y)-z(x)]dy.
\end{equation}
The function $\Omega$ {\big(}as well as the function $\gamma$ from
(\ref{eqn2.2}){\big)} is a finite linear combination of
continuous (multiplicative) characters of the group $U$. It is
important that $\Omega (u) \le 0$ for all $u\in U$, and that the
function $y\mapsto \Omega(u_y)$ is locally constant on $K$.

The Green function $E$ is a non-negative function. If
$\al>{\frac{n}{2}}$, then $E$ is continuous on the whole $K$,
while for $\al={\frac{n}{2}}$ the function $E$ is continuous
except at the origin where it has a logarithmic singularity
\begin{equation*}
%\label{eqn2.7}%
E(x)\le
C_1\left|\,\log\|x\|\,\right|+C_2 \quad {\rm for}\quad \|x\|\le 1
\end{equation*}
($C_1, C_2\ge 0$). This property resembles the property of the
Green function over ${\mathbb R}^2$. As $\|x\|\to\infty$,
$E(x)\le{\rm const}\|x\|^{-2\al/n-1}$ (note a misprint in Ref. 5
where the sign is confused in the expression for the order of
decay of $E$ in the formula (2.25) of Ref. 5; $\|x\|^{\nu-1}$ in
that formula should be replaced with $\|x\|^{-\nu-1}$).

\subsection{Restriction of $\AA$ to the union of balls}
\noindent \looseness1 In order to construct the mixed interaction
measure, we need the Green function $E_\Pi$ of the operator
$\AA_\Pi$. Just as it was explained above for the operator $A_O$,
$\AA_\Pi$ is defined on a function $f\in\DPi$ as the function $\AA
f$  restricted to $\Pi$ (note that $f$ equals zero outside
$\Pi$). It is clear that $\APi$ is symmetric and positive as an
operator on $L_2(\Pi)$. Moreover, $\APi$ is essentially
self-adjoint.

Indeed, let $\AtPi$ be the Friedrichs extension of $\APi$. The
open compact set $\Pi$ is a union $\bigcup\limits_{i=1}^\nu O_i$
of disjoint balls of the same radius, $O_i=\{\,x\in K \vert
\|x-x_i\|\le q^k\,\}$, where $x_i\in K$, $k\in{\mathbb Z}$,
$\|x_i-x_j\|>q^k$ for $i\ne j$. If $y\in K$, and $\|y\|$ is small
enough, then the shift operator $(T_yf)(x)=f(x+y)$ is a unitary
operator on $L_2(\Pi)$. It follows from (\ref{eqn2.6}) that
$\APi$ and $T_y$ commute; then $T_y$ also commute with
$\AtPi$.\cite{Ph} Therefore if $z\in L_2(\Pi)$ is a solution of
the equation $(\APi+m^2)z=f$ where $f\in \DPi$ (i.e., $f$ is
locally constant), then $z$ is locally constant and belongs to the
domain of the operator $\APi$. This means that the positive
definite operator $\APi+m^2$ has a dense range, whence $\APi$ is
essentially self-adjoint (see Ref. 12, Theorem X.26).

In order to write a hyper-singular integral representation for $\APi$, denote $$ R_i=\{\,y\in K
\vert \|y-(x_i-x_j)\|>q^k\ \ {\rm for\ all}\ \ j=1,\ldots,\nu\,\}. $$

\begin{lemma}
Let $x\in O_i$. Then $x-y\in \Pi$ if and only if
$y\not\in R_i$.
\end{lemma}

\begin{proof}
If $y\not\in R_i$, then $\|y-(x_i-x_j)\|\le q^k$ for
some $j$, whence
$$
\|(x-y)-x_j\|=\|(x-x_j)-(y-(x_i-x_j))\|\le q^k
$$
by the ultra-metric property of the absolute value, so that
$x-y\in O_j$.

Conversely, if $y\in R_i$, then for any $j$
$$
\|(x-y)-x_j\|=\|(x-x_j)-(y-(x_i-x_j))\|> q^k.
$$
\end{proof}

Now for $z\in\DPi$ we can rewrite (\ref{eqn2.6}) as follows.
If $x\in O_i$, $y\in R_i$, we have $z(x-y)=0$. Hence
\begin{multline}
\label{eqn2.8} (\AA_\Pi z)(x)=\int_{K\setminus R_i}
\|y\|^{-2\al/n-1} \Omega (u_y) [z(x-y)-z(x)]\,dy \\
 -\ z(x)\int_{R_i}
\|y\|^{-2\al/n-1} \Omega (u_y)\,dy\ ,\quad x\in O_i.
\end{multline}

\begin{proposition}
$(\AA_\Pi+m^2)^{-1}$ is an integral operator on $L_2(\Pi)$ with a
kernel of the form
\begin{equation}
\label{eqn2.9} E_{\Pi}(x,y)= E(x-y) + \Phi (x,y),\qquad x,y\in\Pi,
\end{equation}
where $\Phi(x,y)$ is locally constant in
two variables. The function $E_\Pi$ satisfies the inequality
\begin{equation}
\label{eqn2.10} 0\le E_\Pi(x,y)\le E(x-y), \qquad x,y\in \Pi.
\end{equation}
\end{proposition}

\begin{proof} Define a family of functions $\Psi_{\xi}(x)$ on $\Pi$
($\xi\in\Pi$) setting
\begin{equation}
\label{eqn2.11} \Psi_\xi(x)=\int_{R_i} E(x-y-\xi)\,
\|y\|^{-2\al/n-1}\, \Omega(u_y)\,dy\ ,\quad x\in O_i,\quad
i=1,\ldots,\nu.
\end{equation}

Let $\Phi(x,\xi)$, for each fixed $\xi\in\Pi$, be the solution of
the equation
\begin{equation}
\label{eqn2.12} (\AA_\Pi + m^2)\Phi(\cdot,\xi)=\Psi_\xi.
\end{equation}
It follows from the local constancy of the function $y\mapsto\Omega(u_y)$,
and the fact that $\|y\|>q^k$ for any $y\in R_i$, that $\Psi_\xi(x)$ is
a locally constant function on $\Pi$, uniformly with respect to $\xi$.
Since $\AA_\Pi$ commutes with small shifts, we find that $\Phi(x,\xi)$
is locally constant in $x$, uniformly with respect to $\xi$.

On the other hand, by (\ref{eqn2.11}) $\Psi_\xi(x)$ is also
locally constant in $\xi$ (uniformly with respect to $x\in\Pi$).
Then the uniqueness of a solution of (\ref{eqn2.12}) implies
the local constancy of $\Phi(x,\xi)$ in $\xi$, uniformly with
respect to $x$. Hence $\Phi$ is locally constant in two variables.

Let $f\in\DPi$, $$ v(x)=\int_\Pi E_\Pi(x,y)\,f(y)\,dy. $$ Then
$v=v_1+v_2$, $$ v_1(x)=\int_\Pi E(x-y)\,f(y)\,dy, \qquad
v_2(x)=\int_\Pi \Phi(x,y)\,f(y)\,dy,\quad x\in\Pi\,; $$ $$
\widetilde{v}_1(x)= \begin{cases}
                        v_1(x), & x\in\Pi \\
                        0,      & x\in K\setminus\Pi
                    \end{cases}\ , \qquad
                    x\in K.
$$ Then $\widetilde{v}_1=\widetilde{v}_{11}+\widetilde{v}_{12}$
where $$ \widetilde{v}_{11}(x)= \int_\Pi E(x-y)\,f(y)\,dy, \qquad
x\in K\,; $$ $$ \widetilde{v}_1(x)= \begin{cases}
                        0, & x\in\Pi\,; \\
                        - \int_\Pi E(x-y)\,f(y)\,dy,      &
                        x\not\in\Pi .
                                   \end{cases}
$$ If $x\in O_i$, then by Lemma 2.2
\begin{multline}
\label{eqn2.13} (\AA\, \widetilde{v}_{12})(x)=\int_{R_i}
\|y\|^{-2\al/n-1}\Omega (u_y)\,\widetilde{v}_{12}(x-y)\,dy \\
 =\quad -\int_\Pi f(\xi)\,d\xi\int_{R_i}
E(x-y-\xi)\,\|y\|^{-2\al/n-1}\Omega (u_y)\,dy \\ =\quad -\int_\Pi
\Psi_\xi(x)\,f(\xi)\,d\xi.
\end{multline}
Note that $(\AA+m^2)\,\widetilde{v}_{12}=\AA\,\widetilde{v}_{12}$
on $\Pi$, and  $(\AA+m^2)\,\widetilde{v}_{11}= f$ on $\Pi$, so
that
\begin{equation*}
\left((\AA_\Pi+m^2)\,{v}_{1}\right)(x)=\ f(x) - \int_\Pi
\Psi_\xi(x)\,f(\xi)\,d\xi, \qquad x\in\Pi.
\end{equation*}
Calculating $(\AA_\Pi+m^2)\,{v}_{2}$ we find that $$
\left((\AA_\Pi+m^2)\,{v}\right)(x)\ =\  - \int_\Pi
\Psi_y(x)\,f(y)\,dy, \qquad x\in\Pi ;$$ together with
(\ref{eqn2.13}) this yields the required equality
$(\AA_\Pi+m^2)\,{v}_{2}=f$ on $\Pi$, for any $f\in\DPi$. Since the
kernel (\ref{eqn2.8}) generates a bounded operator on
$L_2(\Pi)$, and $\DPi$ is dense in $L_2(\Pi)$,
(\ref{eqn2.9}) actually is  the Green function.

Let $f(x)\ge 0$ for all $x\in\Pi$. Suppose that $v$ is not
non-negative. Then there exists such $x_0\in\Pi$ that
$$v(x_0)=\min\limits_{x\in\Pi} v(x) < 0.$$ Since $\Omega(u)\le 0$
on $U$, it follows from (\ref{eqn2.8}) that
$\left((\AA_\Pi+m^2)\,v\right)(x_0)<0$. On the other hand,
$\left((\AA_\Pi+m^2)\,v\right)(x_0)=f(x_0)\ge 0$. This
contradiction proves the lower bound in (\ref{eqn2.10}).

In order to prove the upper bound, consider the function
$$w(x)=\int_\Pi \left[
E(x-\xi)-E_\Pi(x,\xi)\right]\,f(\xi)\,d\xi,$$ with $f(\xi)\ge 0$
on $\Pi$, $f\in\DPi$. We find again, that if
$w(x_0)=\min\limits_{x\in\Pi} w(x)<0$, then
$\left((\AA_\Pi+m^2)\,w\right)(x_0)<0$. Denote $$w_1=\int_\Pi
E(x-\xi)\,f(\xi)\,d\xi,\qquad w_2=\int_\Pi
E_\Pi(x,\xi)\,f(\xi)\,d\xi.$$ By (\ref{eqn2.8}), if\ $\,i\,$\
is such that $x_0\in O_i$, then
\begin{multline*}\left((\AA_\Pi+m^2)\,{w}_{1}\right)(x_0)=\int_{K\setminus
R_i} \|y\|^{-2\al/n-1}\,\Omega (u_y)\,[w_1(x_0-y)-w_1(x_0)]\,dy\\
\ -\ w_1(x_0) \int_{R_i} \|y\|^{-2\al/n-1}\,\Omega (u_y)\,dy\ \ +\
\ m^2w_1(x_0) \\ =\
\left((\AA+m^2)\,{w}_{1}\right)(x_0)-\int_{R_i}
\|y\|^{-2\al/n-1}\,\Omega (u_y)\,w_1(x_0-y)\,dy \\ =\ f(x_0)\ -\
\int_{R_i} \|y\|^{-2\al/n-1}\,\Omega (u_y)\,w_1(x_0-y)\,dy,
\end{multline*}
so that $$\left((\AA_\Pi+m^2)\,{w}\right)(x_0)= - \int_{R_i}
\|y\|^{-2\al/n-1}\Omega (u_y)\,w_1(x_0-y)\,dy \ge 0,$$ since
$w_1(x_0-y)\ge 0$ for all $y$, while $\Omega(u_y)\le 0$. We have
come to a contradiction.\bigskip
\end{proof}

Let $\Pi_1$, $\Pi_2$ be open compact subsets of $K$, such that
$\Pi_1 \subset \Pi_2$. If $E_{\Pi_1}$ and $E_{\Pi_2}$ are the
Green functions of the operators $\AA_{\Pi_1}$ and $\AA_{\Pi_2}$
respectively, then
\begin{equation}
\label{eqn2.14} E_{\Pi_1}(x,y)\ \le\ E_{\Pi_2}(x,y)\quad\text{for
all}\quad x,y\in \Pi_1.
\end{equation}
The proof of this inequality is similar to the proof of the upper
bound in (\ref{eqn2.10}).

\medskip
\section{The Wick renormalization}
\noindent \looseness1 In this section we consider (for our
situation) the renormalization procedure known as ``Wick
ordering''. Our purpose here is to define some non-Gaussian
process on $\DCK$ which we could identify with power of the
Gaussian process $\vphi(\cdot)$.

\subsection{Basic notions}
\noindent \looseness1 Following Ref. 13, we will interpret our
free process $\vphi$ as the Gaussian process with mean zero
indexed by the real Hilbert space $\HH$ obtained by completing
$\DK$ with respect to the inner product
$(f,\,g)_{\AA}=\left((\AA+m^2)^{-1}f,\,g\right)$. We fix the
probability space as $(\DCK, \Sigma, \mu_0)$ where $\Sigma$ is the
$\sigma$-algebra generated by cylindrical sets, $\mu_0$ is the
measure defined (via the Minlos theorem) by the characteristic
functional
\begin{equation*}
%\label{eqn3.1}%
\int_{\DCK} e^{i\vphi(f)}\,d\mu_0 (\vphi) =
e^{-\frac{1}{2}(f,\,f)_\AA},\qquad f\in\DK.
\end{equation*}
Note that $\DK$ is a nuclear space.\cite{Bruhat}

Below we will write $L_\rho(\HH)$ instead of $L_\rho(\DCK, \Sigma,
\mu_0)$, and $\Gamma(\HH)$ instead of $L_2(\HH)$. Let
$\Gamma(\HH)_{\le k}$ be the closure in $\Gamma(\HH)$ of the
linear span of all elements $\vphi(f_1)\ldots\vphi(f_l)$, $l\le
k$, and $\Gamma(\HH)_k$ be the orthogonal complement of
$\Gamma(\HH)_{\le k-1}$ in $\Gamma(\HH)_{\le k}$. It is well known
(see e.g. Ref. 13) that
$\Gamma(\HH)=\bigoplus\limits_{k=0}^\infty \Gamma(\HH)_k$, and if
$\psi\in\bigoplus\limits_{l=0}^k \Gamma(\HH)_l$, then
\begin{equation}
\label{eqn3.2} \|\psi\|_\rho\ \le\
(\rho-1)^{k/2}\,\|\psi\|_2\quad\text{for all}\quad \rho\ge 2,
\end{equation}
where the $\rho$ subscript denotes the $L_\rho$-norm.

As usual, the Wick product $\wick{\vphi(f_1)\cdots\vphi(f_k)}$ of
the Gaussian random variables $\vphi(f_1),\ldots,\vphi(f_k)$ is
the orthogonal projection of an element
$\vphi(f_1)\cdots\vphi(f_k)$ to $\Gamma(\HH)_k$. We will write
$\wick{\vphi(f)^k}=\ \, \wick{\vphi(f)\cdots\vphi(f)}$ ($k$
times). This is a Gaussian random variable with mean zero and the
variance $k!\,(f,\,f)_\AA^k$. More generally, if
$f_1,\ldots,f_k,g_1,\ldots,g_k\in\HH$, then
\begin{equation}
\label{eqn3.3} \langle\wick{\vphi(g_1)\cdots\vphi(g_k)}\,
\wick{\vphi(f_1)\cdots\vphi(f_k)}\rangle = \sum\limits_\pi\,
(\,g_{\pi(1)},\,f_1)_\AA\,\ldots\,(\,g_{\pi(k)},\,f_k)_\AA\
\end{equation}
where the sum  is taken over all the permutations of the indices
$1,\ldots,k$.

 In order to define the Wick power $\wick{\vphi^k}$ of the
Gaussian generalized stochastic process, we approximate $\vphi$ by
ordinary random Gaussian functions, to which the above operation
can be applied. The construction of this approximation is specific
for the non-Archimedean case.

Let $B_{\vk}(0)$ be the ball $\{\,x\in K\,\vert\ \|x\|\le q^{-\vk}\,\}$ with the center at the
origin ($\vk=1,2,\ldots$). Denote by $\Delta_{-\vk}(x)$ the indicator of $B_{-\vk}(0)$, and by
$\delta_{\vk}$ delta-like sequence $$\delta_{\vk}=q^\vk \Delta_{-\vk}(x),\qquad x\in K.$$ This
sequence converges to $\delta$ in $\DCK$. The Fourier transform $\widehat\delta_\vk$ is the
indicator of the ball with radius $q^\vk$ centered at the origin. Since $\delta_\vk\in \DK$, the
convolution $$\vphi_\vk (\xi) = (\vphi\ast\delta_\vk)(\xi), \qquad \xi\in K,$$ is an ordinary
locally constant function. On the other hand, by the definition of convolution, for fixed
$\xi$,\ \ $\vphi_\vk (\xi)=\phi\,(\delta_\vk^{(\xi)})$ where
$\delta_\vk^{(\xi)}=\delta_\vk(x-\xi)$. Thus we can write the Wick power $\wick{\vphi_\vk
(\xi)^k}$, and then associate with it a generalized process by the formula
$$\wick{{\vphi_\vk^k}}(g)=\int_K \wick{\vphi_\vk (x)^k}\,g(x)\,dx,\qquad g\in\DK.$$

We can write explicitly~\cite{Si} that
\begin{equation}
\label{eqn3.4} \wick{\vphi_\vk
(x)^k}=\sum\limits_{j=0}^{[\frac{k}{2}]}
\frac{(-1)^jk!}{2^jj!(k-2j)!}\, \vphi_\vk (x)^{k-2j}\,c_\vk^{2j},
\end{equation}
and, conversely,
\begin{equation*}
%\label{eqn3.5}%
{\vphi_\vk (x)^k}=\sum\limits_{j=0}^{[\frac{k}{2}]}
\frac{k!}{2^jj!(k-2j)!}\, \wick{\vphi_\vk
(x)^{k-2j}}\,c_\vk^{2j},
\end{equation*}
where
\begin{equation*}
%\label{eqn3.6}%
c_\vk^2=\int_{\|\xi\|\le q^\vk}
\left(a(\xi)+m^2\right)^{-1}\,d\xi.
\end{equation*}

\begin{proposition}
Let $\al\ge\frac{n}{2}$. Then for any $r\in(1,2)$ and $g\in\DK$
there exist such positive constants $\tau$ and $C$ that
\begin{equation}
\label{eqn3.7} \|\wick{\vphi_{\vk_1}^k}(g)\ -\
\wick{\vphi_{\vk_2}^k}(g)\|_2\ \le\ C\,\|g\|_r\,q^{-\tau\vk},
\end{equation}
where $\vk=\min\{\vk_1, \vk_2\}$.
\end{proposition}
\begin{proof} For a given $r\in(1,2)$ we take $r'=(1-\frac{1}{r})^{-1}$,
$2<r'<\infty$ and $s=\frac{r'}{2}$. Then by the Hausdorff-Young
inequality (see Ref. 3, Theorem 31.22) we have
$\|\,\wh{g}\,\|_{r'}\le\|g\|_r$. Therefore
\begin{equation}
\label{eqn3.8} \|\,\vert\,\wh{g}\,\vert^2\|_s\ \le\ \|g\|_r^2,
\end{equation}
since
$\|\,\vert\,\wh{g}\,\vert^2\|_s=\|\,\vert\,\wh{g}\,\vert^2\|_{r'/2}=
\|\,\vert\,\wh{g}\,\vert\,\|_{r'}^2$.

It follows from (\ref{eqn3.3}) and (\ref{eqn3.4}) that for
$\vk_1\ge\vk_2$
\begin{equation}
\label{eqn3.9} (k!)^{-1}\|\wick{\vphi_{\vk_1}^k}(g)\ -\
\wick{\vphi_{\vk_2}^k}(g)\|_2^2\ =\ \left(g,\,E_{\vk_1}^{\,k}\ast
g \right)_{L_2(K)}-\left(g,\,E_{\vk_2}^{\,k}\ast g
\right)_{L_2(K)},
\end{equation}
where
\begin{equation*}
%\label{eqn3.10}%
E_{\vk_i}(x)=\int_K \delta_{\vk_i}(x-y)\,E(y)\,dy.
\end{equation*}
The derivation of (\ref{eqn3.9}) is a straightforward
calculation based on the identity\par\nobreak\noindent
$\delta_{\vk_1}*\delta_{\vk_2}=\delta_{\min\{\vk_1,\vk_2\}}$.

From (\ref{eqn3.8}) and (\ref{eqn3.9}) taking in account
the Plancherel equality we get
\begin{multline*}(k!)^{-1}\|\wick{\vphi_{\vk_1}^k}(g)\ -\
\wick{\vphi_{\vk_2}^k}(g)\|_2^2=\|\,\vert\,\wh{g}\,\vert^2\,
(\wh{E}_{\vk_1}^{\ast k}-\wh{E}_{\vk_2}^{\ast k})\,\|_{L_1(K)}\
\le
\\ \|\,\vert\,\wh{g}\,\vert^2\|_s\, \|\,\wh{E}_{\vk_1}^{\ast
k}-\wh{E}_{\vk_2}^{\ast k}\,\|_{s'}\ \le\ \|\,g\,\|_r^2\,
\|\,\wh{E}_{\vk_1}^{\ast k}-\wh{E}_{\vk_2}^{\ast
k}\,\|_{s'},\qquad
\end{multline*}
where $s'>1$, namely $s'^{-1}=1-1/s$.

Now it is sufficient to show that the estimate
$$\|\,\wh{E}_{\vk_1}^{\ast k}-\wh{E}_{\vk_2}^{\ast k}\,\|_{s'}\le
O(q^{-\tau\vk})$$ holds for some $\tau>0$. But
\begin{multline*}
\wh{E}_{\vk_1}^{\ast k}-\wh{E}_{\vk_2}^{\ast k}\\ =
(\wh{E}_{\vk_1}-\wh{E}_{\vk_2})\ast\wh{E}_{\vk_2}\ast\cdots\ast
\wh{E}_{\vk_2}+\wh{E}_{\vk_1}\ast(\wh{E}_{\vk_1}-\wh{E}_{\vk_2})
\ast\wh{E}_{\vk_2}\ast\cdots\ast \wh{E}_{\vk_2}+\cdots \\ \cdots +
\wh{E}_{\vk_1}\ast\cdots\ast\wh{E}_{\vk_1}(\wh{E}_{\vk_1}-
\wh{E}_{\vk_2}).
\end{multline*}
From (\ref{eqn2.5}) and the definition of $E_{\vk_i}$ we find
that for $\rho>1$ the norm $\|\wh{E}_{\vk_i}\|_\rho$ is uniformly
bounded with respect to $\vk=\min\{\vk_1,\vk_2\}$. Then by the
Young inequality (see Ref. 3, Theorem 31.45) for
$\rho\in(1,1+\frac{1}{k})$ there exists a positive constant, such
that $$\|\,\wh{E}_{\vk_1}^{\ast k}-\wh{E}_{\vk_2}^{\ast
k}\,\|_{s'} \le
\text{const}\|\,\wh{E}_{\vk_1}-\wh{E}_{\vk_2}\,\|_{\rho}.$$
Therefore the only thing we have to show is the estimate
\begin{equation*}
\|\,\wh{E}_{\vk_1}-\wh{E}_{\vk_2}\,\|_{\rho}\ \le\
O(q^{-\tau\vk}),
\end{equation*}
but it follows directly from (2.4) and the inequality
$$\|\,\wh{E}_{\vk_1}-\wh{E}_{\vk_2}\,\|_{\rho}\le \int_{\|\xi\|\ge
q^\vk}\left(a(\xi)+m^2\right)^{-\rho}\,d\xi.$$
\end{proof}

The last proposition implies that for any function $g\in\DK$ the
sequence $\{\wick{\vphi_\vk^k}(g)\}$  converges, as
$\vk\to\infty$, to some random variable $\wick{\vphi^k}(g)$.
Moreover, the limit $\wick{\vphi^k}(g)$ belongs to
$\Gamma(\HH)_k$, since $\wick{\vphi_\vk^k}(g)$ is in
$\Gamma(\HH)_k$ for any $\vk$.

 It follows from (\ref{eqn3.9}) that
\begin{equation*}
%\label{eqn3.12}%
(k!)^{-1}\|\wick{\vphi^k}(g)\ -\
\wick{\vphi_{\vk}^k}(g)\|_2^2\ =\ \left(g,\,E^{\,k}\ast g
\right)_{L_2(K)}-\left(g,\,E_{\vk}^{\,k}\ast g \right)_{L_2(K)}.
\end{equation*}

Passing to the limit in relation (\ref{eqn3.7}) yields
\begin{equation}
\label{eqn3.13} \|\wick{\vphi^k}(g)\ -\
\wick{\vphi_{\vk}^k}(g)\|_2\ \le\ C\,\|g\|_r\,q^{-\tau\vk},
\end{equation}
where $C$ and $\tau$ are some positive constants, and $1<r<2$.

Since $\wick{\vphi^k}(g)$ belongs to $\Gamma(\HH)_k$, we can apply
the inequality (\ref{eqn3.2}). Taking in account
(\ref{eqn3.13}) we get the following proposition.
\medskip
\begin{proposition} If $\al\ge\frac{n}{2}$, then for any
$r\in(1,2)$, $g\in\DK$, and $\rho\ge 2$ there exist such positive
constants $\tau$ and $C$ that
\begin{equation*}
%\label{eqn3.14}%
\|\wick{\vphi^k}(g)\ -\
\wick{\vphi_{\vk}^k}(g)\|_{\rho}\ \le\
(\rho-1)^{k/2}\,C\,\|g\|_r\,q^{-\tau\vk}.
\end{equation*}
\end{proposition}

\subsection{The Wick polynomials}
\noindent \looseness1 Let $g\in\DK$. For any polynomial
$P(X)=a_sX^s+\cdots+a_1X+a_0$ we define $\wick{P(\vphi)}(g)=\int_K
g(x)\wick{P\left(\vphi(x)\right)}\,dx$ as
\begin{equation*}
%\label{eqn3.15}%
a_s\wick{\vphi^s}(g)+\cdots+a_2\wick{\vphi^2}(g)+
a_1\vphi(g)+a_0.
\end{equation*}
Similarly we define the ``smoothed'' polynomial
$\wick{P(\vphi_{\vk})}(g)$. In what follows we will assume that
$a_s>0$ and $g(x)>0$ for any $x\in \text{supp}\,g$.

It is quite obvious that if $s=\text{deg}\,P$ is an even number, then the polynomial $P$ is
bounded from below. So it is natural to expect that under our assumptions the integral
\begin{equation}
\label{eqn3.16}
\int_{\DCK}
\exp\left(-\wick{P(\vphi)}(g)\right)\,d\mu_0(\vphi)
\end{equation}
should converge. However, the Wick renormalization procedure
usually causes the loss of the semiboundedness. It will take us
two additional steps to show convergence of the integral
(\ref{eqn3.16}).

Denote $$B=\max\limits_{0\le j\le s}|a_j|,$$ and
$$D=a_s\,\|g\|_{L_1(K)}\left(1+\max\limits_{0\le j\le s-1} \left(\left|\frac{a_j}{a_s}\right|
+1\right)^{s/(s-j)}\right).$$

\begin{proposition} Let $s=\text{deg}\,P$ be even; then  there exists such a positive constant,
that for any $\vk$
\begin{equation*}
%\label{eqn3.17}%
\wick{P(\vphi_\vk)}(g)\ge -\text{\rm
const}\,D\,\vk^{s/2}.
\end{equation*}
\end{proposition}
\begin{proof} Using the relation (\ref{eqn3.4}) we can rewrite
$\wick{P\left(\vphi_\vk(x)\right)}$ as
\begin{multline*}
\wick{P\left( \vphi_\vk(x)\right)} =  \sum\limits_{j=0}^{s} a_j
\sum\limits_{l=0}^{[\frac{j}{2}]} \frac{j!(-1)^l}{l!(j-2l)!} \vphi_\vk(x)^{j-2l}(c_\vk^2)^l \\
=\quad a_s\left\{\vphi_\vk(x)^s+ \sum\limits_{j=0}^{s-1}
\frac{a_j}{a_s} \sum\limits_{l=0}^{[\frac{j}{2}]}
\frac{j!(-1)^l}{l!(j-2l)!} \vphi_\vk(x)^{j-2l}(c_\vk^2)^l \right.
\\ \left. +\quad \sum\limits_{k=1}^{s/2}
\frac{s!(-1)^k}{k!(s-2k)!} \vphi_\vk(x)^{s-2k}(c_\vk^2)^k\right\}.
\end{multline*}
Let ${\mathcal N}_1$ denote the number of terms in the first sum, and ${\mathcal N}_2$ the
number of terms in the second one; ${\mathcal N}={\mathcal N}_1+{\mathcal N}_2$. Then we have
\begin{multline}
\label{eqn3.18} \wick{P\left(\vphi_\vk(x)\right)} = a_s\sum\limits_{j,l}\left({\mathcal
N}^{-1}\vphi_\vk(x)^s + b_{j,l}\frac{a_j}{a_s} \vphi_\vk(x)^{j-2l} (c_\vk^2)^l\right) \\ +
a_s\sum\limits_{k}\left({\mathcal N}^{-1}\vphi_\vk(x)^s + d_{k}\vphi_\vk(x)^{s-2k}
(c_\vk^2)^k\right),
\end{multline}
where $b_{j,l}$ and $d_k$ are some coefficients.

Elementary computations show that for even $s$ and $j<s$ the
inequality $X^s-c_jX^j\ge-|c_j|^{s/(s-j)}$ holds for all $X$.
Therefore, each term in the first sum of the right-hand side of
(\ref{eqn3.18}) is bounded from below by the value $$-\,C_1
\left|\frac{a_j}{a_s}\right|^{s/(s-j+2l)}\,(c_\vk^2)^{sl/(s-j+2l)}
\ge -\,C_1\left(\left|\frac{a_j}{a_s}\right|+1\right)^{s/(s-j)}\,
(c_\vk^2)^{s/2},$$ since $\frac{2l}{s-j+2l}<1$. By the same
reasoning, each term in the second sum is bounded from below by
the value $$-\,C_2\,(c_\vk^2)^{s/2}.$$ (Here $C_1$ and $C_2$
denote some positive constants.) Thus there exists such a positive
constant that $$\wick{P\left(\vphi_\vk(x)\right)}\ge
-\,\text{const}\,a_s\left(1+\max\limits_{0\le j\le s-1}
\left(\left|\frac{a_j}{a_s}\right|+1\right)^{s/(s-j)}\right)\,(c_\vk^2)^{s/2}.$$
From here and Lemma~2.1 we get
\begin{multline*}
 \wick{P(\vphi_\vk)}(g)= \int_K
\wick{P\left(\vphi_\vk(x)\right)}\,g(x)\,dx\ge
\\
 -\,\text{const}\,a_s\,\|g\|_{L_1(K)}
\left(1+\max\limits_{0\le j\le s-1}
\left(\left|\frac{a_j}{a_s}\right|+1\right)^{s/(s-j)}\right)\,(c_\vk^2)^{s/2}
\\ \ge -\,\text{const}\,D\,\vk^{s/2}.
\end{multline*}
\end{proof}

It follows directly from the Proposition~3.1 and the definitions of $\wick{P(\vphi)}(g)$ and
$\wick{P(\vphi_\vk)}(g)$ that for any $r\in(1,2)$ there exist such $C>0$ and $\tau>0$ that
\begin{equation}
\label{eqn3.19} \|\wick{P(\vphi)}(g)-\wick{P(\vphi_\vk)}(g)\|_\rho
\le C\,(\rho-1)^{s/2}\,B\,\|g\|_r\,q^{-\tau\vk}
\end{equation}
for any $\rho\ge 2$.

\begin{proposition} Let $P$ be as in the previous
proposition. Then there exist such positive constants $b$ and
$\gamma$, that
\begin{equation}
\label{eqn3.20} \mu_0\{\vphi\ \vert\quad \wick{P(\vphi)}(g) \le -b\vk^{s/2}\,\} \le
e^{-q^{\gamma\vk}}
\end{equation}
if $\vk$ is large enough.
\end{proposition}
\begin{proof} First, we choose a constant $b$ such that the
estimate $\wick{P(\vphi_\vk)}(g)\ge -\frac{b}{2}\vk^{s/2}$ holds;
Proposition~3.3 allows us to make this choice. Then we take
$\vk_0$ such that $\frac{b}{2}\vk_0^{s/2}>1$, so that
$$\wick{P(\vphi_{\vk})}(g)\ge 1-b\,\vk^{s/2}\quad\text{\ for\ any}\quad \vk\ge \vk_0.$$

If $\wick{P(\vphi)}(g)\le -\,b\,\vk^{s/2}$, then for all $\vk\ge \vk_0$ we have
$$|\wick{P(\vphi)}(g)-\wick{P(\vphi_\vk)}(g)|\ge 1$$, and, therefore
\begin{multline*}
\mu_0\{\vphi\ \vert\quad \wick{P(\vphi)}(g)\le -b\,\vk^{s/2}\} \le \mu_0\{\vphi\ \vert\quad
|\wick{P(\vphi)}(g)-\wick{P(\vphi_\vk)}(g)|\ge 1\} \\ \le \int_{\DCK}
|\wick{P(\vphi)}(g)-\wick{P(\vphi_\vk)}(g)|^\rho\,d\mu_0(\vphi) =
\|\wick{P(\vphi)}(g)-\wick{P(\vphi_\vk)}(g)\|_\rho^\rho
\end{multline*}
for all $\rho\ge2$. From inequality (\ref{eqn3.19}) it follows
that there exists such a constant $R$, which does not depend on
$\vk$, that
$$\|\wick{P(\vphi)}(g)-\wick{P(\vphi_\vk)}(g)\|_\rho^\rho\le
(\rho-1)^{s\rho/2}\,R^\rho\,q^{-\tau\vk\rho},$$ where $\tau>0$. Then we take $\rho$ depending on
$\vk$; namely, we put $\rho=q^{2\tau\vk/3}$. We have then $$(\rho-1)^{s\rho/2}\le \rho^{s\rho/2}
\le q^{\tau\vk\rho/3}.$$ Moreover, it is clear that for all sufficiently large $\vk$'s we have
$R^\rho\le q^{\tau\vk\rho/3}$, and $e\le q^{\tau\vk\rho/3}$. Thus,
$$R^\rho(\rho-1)^{s\rho/2}\,q^{-\tau\vk\rho}\le q^{-\tau\vk\rho/3} \le e^{-\rho} =
e^{q^{-2\tau\vk\rho/3s}} = e^{-q^{\gamma\vk}},\quad\text{ where } \gamma=\frac{2\tau}{3s}.$$
\end{proof}

Now everything is prepared for proving the main result of this
section which is formulated in the following theorem.
\medskip
\begin{theorem} Let $P$ be a polynomial bounded from below (that is, {\rm deg}$\,P=2s$),
and $g$ be a nonnegative function from $\DK$. Then
\begin{equation}
\label{eqn3.21} \exp\left(-\int_K g(x)\,\wick{P(\vphi(x))}\,dx
\right)\in\bigcap\limits_{\rho<\infty} L_\rho(\DCK,d\mu_0).
\end{equation}
\end{theorem}
\begin{proof} Let $f$ be a real-valued function on a probability space
$(M,\Sigma,\mu)$ and $$m_f(x)=\mu\{\xi\ \vert\quad f(\xi)\ge x\}.$$ Next let $F$ be a bounded
positive function from $C^1(\mathbb{R})$. Then
\begin{multline*}
\int F(f(g))\,d\mu = \int_{-\infty}^{+\infty} F(x)\,dm_f\quad \text{(Stieltjes' integral)} \\ =\
-\,F(-\infty)+\int_{-\infty}^{+\infty} m_f(x)\,F'(x)\,dx.\qquad
\end{multline*}
By the monotone convergence theorem we have $$\int
e^{f(\xi)}\,d\mu = \int_{-\infty}^{+\infty} e^x\,m_f(x)\,dx,$$
where both sides of the formula can turn into infinity only
simultaneously. If the function $f$ is such that $$\mu\,\{\,\xi\
\vert\quad -f(\xi)\ge b\vk^{s/2}\,\} \le e^{-q^{\gamma\vk}}$$ for
all $\vk\ge \vk_0$, then $$\int e^{-f(\xi)}\,d\mu \le
e^{b\vk^{\,s/2}}+\int_{b\vk^{\,s/2}}^{\infty}
e^x\,\exp\left(-q^{\gamma(x/b)^{2/s}}\right)\,dx < \infty .$$ From
here and the Proposition~3.4 we conclude, that the integral
$$\int_{\DCK} \exp\left(-\wick{P(\vphi)}(g)\right)\,d\mu_0(\vphi)$$
converges. Thus, we have proved (\ref{eqn3.21}) for the case
of $\rho=1$. But
$$\left\|\exp\left(-\wick{P(\vphi)}(g)\right)\right\|_\rho =
\left\|\exp\left(-\wick{P(\vphi)}(\rho
g)\right)\right\|_1^{1/\rho},$$ which proves (\ref{eqn3.21}).
\end{proof}

Now we can state that the expression
\begin{equation}
\label{eqn3.22}
d\mu_g(\vphi)={\frac{\exp\left(-\wick{P(\vphi)}(g)\right)\,d\mu_0}{
\int\exp\left(-\wick{P(\vphi)}(g)\right)\,d\mu_0}}
\end{equation}
defines a probability measure on the space $\DCK$.

\medskip
\section{Concentrated measure and Schwinger functions}
\noindent \looseness1 In this section we introduce a
``concentrated'' measure $\dmuPi$ on the space $\DCPi$
corresponding to the operator $\APi$ defined in section~2. We show
that the expression similar to that in~(\ref{eqn3.22}) with
$d\mu_0$ replaced by $d\mu_0^{\Pi}$ makes sense and defines a
probability measure on $\DCPi$. Next we consider the natural
counterparts of the ``semi-Dirichlet'' Schwinger functions
corresponding to a bounded region and examine their behaviour as
the region increases.

Let $\Pi$ be a union of balls of the same radius $q^k$ described
in subsection~2.3, and $\APi$ be the restriction of the operator
$\AA$ to that union. In this section we accept a little more
verbose notations, representing the union in form
$\Pi=\bigcup_{i=1}^{\nu} O_k(x_i)$, where the common radius of
balls is mentioned explicitly (subsection~2.3 pointed the ball
$O_k(x_i)$ by saying simply $O_i$).

\subsection{The measures}
\noindent \looseness1 The properties of the Green function
$E_{\Pi}$ of the operator $\APi$ studied in detail in
subsection~2.3 allow us to define on the space $\DCPi$ a
generalized stochastic process $\vphi$ with mean zero and
covariance of form
\begin{equation*}
%\label{eqn4:1}%
\left(f,(\APi+m^2)^{-1}g\right) =\int_{K}
f(x)\,E_{\Pi}(x,y)\,g(y)\,dy\,dx,\quad f,g\in\DPi.
\end{equation*}
The corresponding Gaussian measure on $\DCPi$ will be denoted by
$\dmuPi$. At the same time we can consider the restriction of the
measure $d\mu_0$ to $\DCPi$; it will be denoted by $d\mu_0$ as
well.

Thus we have two different measures on $\DCPi$, and therefore two
kinds of the Wick renormalization for random variables on $\DCPi$,
such as the generalized process $\vphi$. Since $\vphi$ is 
Gaussian with respect to each of the two measures,
one can state (just as in the case of ${\mathbb R}^2$; see Ref. 13) that
\begin{equation}
\label{eqn4:2} \wick{\vphi
(x)^r}=\sum\limits_{j=0}^{[\frac{r}{2}]}
\frac{r!}{2^jj!(r-2j)!}\,\wick{\vphi
(x)^{r-2j}}_{\Pi}\,\left(\Phi (x,x)\right)^j,
\end{equation}
for any natural number $r$, where $\wick{\cdot}$ and
$\wick{\cdot}_{\Pi}$ denote the Wick renormalization with respect
to $d\mu_0$ and $\dmuPi$, respectively. From Proposition~2.1 we find that
$\Phi (x,x)$ is a bounded, nonpositive, and locally constant
function in $x$. The formula~(\ref{eqn4:2}) implies that the
polynomial $\wick{P(\vphi (x))}=\sum_{j=0}^{s} a_j\wick{\vphi
(x)^j}$, whose degree $s$ is even, can be represented as a similar
polynomial $\wick{P'(\vphi (x))}_{\Pi}=\sum_{j=0}^{s}
a_j'\wick{\vphi (x)^j}_{\Pi}$ of the same degree. Moreover, if
$a_s>0$, then $a_s'>0$; in other words, if the polynomial
$\wick{P(\vphi (x))}$ is bounded below, then so is the polynomial
$\wick{P'(\vphi (x))}_{\Pi}$. Therefore, for proving the
convergence of the integral
$$\int_{\DCPi}
e^{-\wick{P(\vphi)}(g)}\,\dmuPi,$$%
where $g$ is a nonnegative function from $\DPi$, $P$ is the
semibounded polynomial, and $\wick{P(\vphi)}(g)=\int_{K}
\wick{P(\vphi (x))}\allowbreak\,g(x)\,dx$, it is sufficient to
show (under the same conditions) the convergence of
\begin{equation*}
%\label{eqn4:3}%
\int_{\DCPi} e^{-\wick{P(\vphi)}_{\Pi}(g)}\,\dmuPi.
\end{equation*}
The convergence of the latter integral can be proved in
essentially the same way as it was done for $\int_{\DPi}
e^{-\wick{P(\vphi)}(g)}\,d\mu_0$ in the previous section. Together
with the formula (\ref{eqn4:2}) this implies that the
expression
\begin{equation}
\label{eqn4:4}
d\mu_g^{\Pi}(\vphi)={\frac{\exp\left(-\wick{P(\vphi)}(g)\right)\,\dmuPi}{
\int\exp\left(-\wick{P(\vphi)}(g)\right)\,\dmuPi}}
\end{equation}
defines a probability measure on the space $\DCPi$.

Concluding this subsection we define a function
\begin{equation}
\label{eqn4:5}%
S_g^{(\Pi)}(h_1,\ldots,h_r)= Z_{\Pi}^{-1}\int
\vphi(h_1)\cdots\vphi(h_r)\,e^{-\wick{P(\vphi)}(g)}\,\dmuPi,
\end{equation}
where $\{\,h_1,\ldots,h_r\}$ is an  arbitrary collection of
functions from $\DPi$, and $Z_{\Pi} = \int
e^{-\wick{P(\vphi)}(g)}\,\dmuPi$. The function
$S_g^{(\Pi)}(h_1,\ldots,h_r)$ is said to be ``$\,${\it the $r$-point
Schwinger function} of a mixed state corresponding to the region
$\Pi$.'' In what follows we show that if $h_i\ge 0$, then the
Schwinger function is nonnegative, and it grows as the region
$\Pi$ increases.

\subsection{Lattice approximation}
\noindent \looseness1 Let $\Pi$ be the union of balls defined at
the beginning of the section. For any integer $l$, such that $l\le
k$, we can write the following identity based on the geometric
features of non-Archimedean fields (see Ref. 15):
$$\Pi=\bigcup_{i=1}^{\eta} O_l(x_i),$$ where $O_l(x_i)$ denotes
a ball $\left\{\,x\in K\ \vert\quad \|x-x_i\|\le q^{l}\,\right\}$
of radius $q^l$, and $\eta=\eta(l)\ge \nu$ is the number of the balls
$O_l(x_i)$ in the union $\Pi$. The balls are disjoint, and their
centers are more than $q^l$ apart from each over, that is
$\|x_i-x_j\|\ge q^l$ for $i\ne j$. This means that we can
decompose $\Pi$ into the union of arbitrarily small parts keeping
general structure of the union unchanged, so that all results of
subsection~2.3 remain valid. Then let
$$e_i^{(l)}(x)=q^{-l/2}\Delta_{O_l(x_i)} (x),$$ where $\Delta
_{O_l(x_i)}$ is the indicator of the ball $O_l(x_i)$. The
collection $\{ e_i^{(l)}\}_{i=1}^{\eta}$ forms an orthonormal
basis in the space $\DlPi$, which consists of all functions from
$\DlPi$ with the following local constancy property:  for any
$f\in \DlPi$ we have $f(x+x')=f(x)$ if $\|x'\| \le q^l$. Note that
$\DlPi$ is a space of the finite dimension $\eta(l)$. Applying the
formula (\ref{eqn2.8}) to the function $z=e_i^{(l)}$ we easily
get
\begin{equation*}
%\label{eqn4.1}%
\left(\APi\,e_i^{(l)}\right)(x)=\begin{cases}
                        -q^{-l/2}\int\limits_{K\backslash O_l(x_i)}\left\|
                        y\right\| ^{-2\alpha
/n-1}\Omega (u_y)dy, & x\in O_l(x_i); \\
                        q^{-l/2}\int\limits_{\left\|(x_i-x_j)-y\right\| \leq q^l}\left\|
y\right\| ^{-2\alpha /n-1}\Omega (u_y)dy, & x\in O_l(x_j),\;i\neq j.
                    \end{cases}
\end{equation*}
It follows from here that the space $\DlPi$ is invariant with respect to the operator $\APi $,
and so it is with respect to the operators $(\APi +m^2)$ and $(\APi +m^2)^{-1}$. Moreover, the
operator $(\APi +m^2)^{-1}$ acts on $\DlPi$ as a positive operator, since it is a restriction of
the positive operator $(\AA+m^2)^{-1}$ to a subspace.

Now consider a matrix $M^{(\Pi,l)}$ with the elements of form
\begin{equation*}
M_{ij}^{(\Pi,\,l)}=(e_i^{(l)},(\APi +m^2)^{-1}e_j^{(l)}),\quad
i,j=1,\ldots,\eta.
\end{equation*}
It is easy to see that this matrix is symmetric and positive
definite. Furthermore, since the integral kernel of $(\APi
+m^2)^{-1}$ is a nonnegative function (see Proposition~2.1), all
the elements of $M^{(\Pi,\,l)}$ are nonnegative. The elements of
its inverse $N^{(\Pi,\,l)}$ have the form
\begin{equation*}
%\label{eqn4.3}%
N_{ij}^{(\Pi,\,l)}=(e_i^{(l)},(\APi +m^2)e_j^{(l)}),
\end{equation*}
and we can express them as
\begin{equation}
\label{eqn4.4}%
N_{ij}^{(\Pi,\,l)}=\left\{
\begin{array}{cc}
\begin{array}{cc}
-\int\limits_{K\backslash O_l(x_i)}\left\| y\right\| ^{-2\alpha /n-1}\Omega (u_y)dy, &
\end{array}
\quad \quad  & i=j; \\ \int\limits_{\left\| (x_i-x_j)-y\right\| \leq q^l}\left\| y\right\|
^{-2\alpha /n-1}\Omega (u_y)dy, & \;i\neq j.
\end{array}
\right.
\end{equation}
Since $\Omega(u)\le 0$ for all $u\in U$, we conclude that
\begin{description}
\item[(a)] $N_{ii}^{(\Pi,\,l)}\ge 0$,\qquad $i=1,\ldots,\eta$;%
\smallskip\noindent%
\item[(b)] $N_{ij}^{(\Pi,\,l)}\le 0$,\qquad $i\ne j,\quad
i,j=1,\ldots,\eta$.
\end{description}
\medskip

There is one more important fact, which follows directly from
(\ref{eqn4.4}). Suppose that we added a number of new balls of
the radius $q^k$ to the union $\Pi$; denote that extended union by
$\Pi'$. Again, for any $l\le k$ we can represent $\Pi'$ as a union
of balls of radius $q^l$; moreover, once we have such a
representation for $\Pi$, we can choose the representation for
$\Pi'$ to be exactly the same plus some additional balls of radius
$q^l$. The corresponding matrix $N^{(\Pi',\,l)}$ will then consist
of all elements of $N^{(\Pi,\,l)}$ plus additional ones. More
precisely, if $\Pi\subset\Pi'$, and $i,j\in\Pi$, then
\begin{equation}
\label{eqn4.5} N_{ij}^{(\Pi,\,l)}=N_{ij}^{(\Pi',\,l)},
\end{equation}
because the matrix elements have exactly the same form, according
to (\ref{eqn4.4}). (Here and below $i\in\Pi$ means that
$O_i\subset\Pi$.)

Next we define a Gaussian stochastic process $\vphi_\delta$ on $\DlPi$ 
with the covariance
\begin{equation*}
%\label{eqn4.6}%
\langle\,\vphi_\delta(e_i^{(l)})\,\vphi_\delta(e_j^{(l)})\,
\rangle=M_{ij}^{(\Pi,\,l)},
\end{equation*}
where $\delta=q^l$. The Gaussian random variables
$\vphi_\delta(e_i)$ form a jointly-Gaussian collection with the
characteristic function
\begin{equation*}
%\label{eqn4.7}%
c(t_1,\ldots,t_{\eta})=\
\exp\left(-\frac{1}{2}\sum_{i,j=1}^{\eta}M_{ij}^{(\Pi,\,l)}t_it_j\right),
\end{equation*}
where $t_l\in {\mathbb R},\ l=\overline{1,\eta}$. The
corresponding Gaussian measure on the space 
${\mathcal D}'_l(\Pi )\cong \DlPi$, which
is essentially a measure on ${\mathbb{R}}^{\eta}$, will be denoted
by $d\mu_{\delta}^{\Pi}$. It is a well known fact (see
e.g. Ref. 13), that if $F$ is a function on ${\mathbb R}^\eta$, then
\begin{multline}
\label{eqn4.8} \int_{{\mathcal D}'_l(\Pi )}
F\left(\vphid(e_1^{(l)}),\ldots,\vphid(e_\eta^{(l)})\right)\,
d\mu_{\delta}^{\Pi}
(\vphid)\\=\
(2\pi)^{-\eta /2}\left(\text{det}M^{(\Pi,\,l)}\right)^{-1/2}\,\int_{{\mathbb
R}^\eta}
F(t_1,\ldots,t_\eta)\,e^{-{\frac{1}{2}}\sum\limits_{i,j=1}^\eta
N_{ij}^{(\Pi,\,l)}t_it_j}\,d^\eta t.
\end{multline}

For any function $g\in \DPi$ we put
\begin{equation*}
%\label{eqn4.9}%
\vphid(g)=\sum_{i=1}^{\eta} g(x_i)\vphid(e_i^{(l)}),
\end{equation*}
where $x_i$ is the center of the ball $O_i$. The random variable
$\vphid(g)$ can be considered as defined on the probability space
$(\DCPi,d\mu_0)$. This allows us to define its Wick powers by the
formula
\begin{equation*}
%\label{eqn4.10}%
\wick{\vphid^r(g)}=\sum_{i=1}^{\eta}
g(x_i)\wick{\vphid^r(e_i^{(l)})},
\end{equation*}
where the renormalization in the right-hand side is taken with
respect to the free measure $d\mu _0$. Since $\vphid(e_i)$ is an
ordinary random variable, there are no difficulties with such a
definition.

For an arbitrary polynomial
$P(X)=a_sX^s+a_{s-1}X^{s-1}+\cdots+a_1X+a_0$ and any function
$g\in\DPi$ we define
\begin{equation*}
%\label{eqn4.11}%
\wick{P(\vphid)}(g)=\sum_{j=0}^s\sum_{i=1}^\eta
a_j\wick{\vphid^j(e_i^{(l)})}g(x_i).
\end{equation*}
Finally, we define the function
\begin{equation}
\label{eqn4.12} \SchdgPi\,(h_1,\ldots,h_r)=Z_{\Pi}^{-1}\int
\vphid(h_1)\cdots\vphid(h_r)\,e^{-\wick{P(\vphid)}(g)}\,d\mu_{\delta}^\Pi,
\end{equation}
where $Z_{\Pi} = \int
e^{-\wick{P(\vphid)}(g)}\,d\mu_{\delta}^\Pi$, the polynomial $P$
is bounded from below (i.e., deg$\,P=s$ is even), $g$ is a
nonnegative function from $\DPi$, and $h_i\in\DPi$,
$i=\overline{1,r}$. The function $\SchdgPi(h_1,\ldots,h_r)$ can be
regarded as the lattice approximation for the $r$-point Schwinger
function $S_{g}^{(\Pi)}(h_1,\ldots,h_r)$  defined above. Note that
for $g\in\DlPi$
$$\vphid(g)=\vphi(g)\qquad\text{and}\qquad \wick{\vphid^k}(g)=\
\wick{{\vphi}^k}(g)\quad\text{for\ any}\ k\in{\mathbb Z}.$$
Moreover, if $h_i\in\DlPi$ then the integral in the right-hand
side of (\ref{eqn4.12}) coincides with  the integral $$\int
\vphi(h_1)\cdots\vphi(h_r)\,e^{-\wick{P(\vphi)}(g)}\,d\mu_0$$ by
the definition of the cylindrical measure $d\mu_0$. Since for any
fixed function from $\DPi$ there exists such an integer $k_0$,
that for all $k\ge k_0$ the function belongs to $\DlPi$, for all
sufficiently small $\delta$'s the Schwinger function and its
lattice approximations are the same thing. Thus, all facts proven
to be true for those approximations with arbitrary $\delta$,
automatically hold for the Schwinger function itself.

\subsection{Griffiths inequalities}
\noindent \looseness1 The Griffiths inequalities are standard
correlation inequalities, which are usually used for proving the
properties of Schwinger functions. To formulate them we will need
some more notions~\cite{Si}.

\begin{definition}
A polynomial $Q$ is said to be even, if it of form
$$Q(X)=a_sX^s+a_{s-2}X^{s-2}+\cdots+a_2X^2+a_0,$$
where $s$ is an even number.
\end{definition}

\begin{definition} A probability measure $\mu$ on
${\mathbb{R}}^r$ is said to correspond to an even Ising
ferromagnet, if it is of form $$\mu = Z^{-1}\exp\left(-\sum
b_{i,j}\,x_ix_j\right)\,d\nu_1\,\ldots\,d\nu_r,$$ where $Z$ is a
normalization constant, $b_{i,j}\le 0$ for $i\ne j$,
$d\nu_i=\exp(\lambda_ix_i)\,d\nu_i'$, $\lambda_i\ge 0$, and
$d\nu_i'$ is a measure which is invariant under reflection, that
is $d\nu_i'(\xi)=d\nu_i'(-\xi)$.
\end{definition}

\smallskip

The fact stated in the following lemma follows directly from the
definitions above and the definition of the measure
$d\mu_{\delta}^{\Pi}$. (Recall here that the space $\DClPi$ is
essentially ${\mathbb{R}}^{\eta}$, so the measure
$d\mu_{\delta}^{\Pi}$ is a Gaussian measure on
${\mathbb{R}}^{\eta}$.)

\begin{lemma} If the polynomial $P$ is of form $P(X)=Q(X)-\lambda X$, where
$Q(X)$ is an even polynomial, $g$ is a nonnegative function from
$\DPi$, and $\lambda$ is a nonnegative constant, then the measure
\begin{multline*}d\nu_{\delta,g}^{\Pi} = Z_{\Pi}^{-1}e^{-\wick{P(\vphid)}(g)}\,
d\mu_{\delta}^{\Pi} \\ = Z_{\Pi}^{-1}
\exp\left(-\frac{1}2\sum_{i,\,{i'}\in\Pi}
N_{i,\,i'}^{(\Pi,\,l)}t_it_{i'}\right)\prod_{j\in\Pi}
\left(e^{-\wick{P(t_j)}(g)}\,dt_j\right)
\end{multline*}
corresponds to an even Ising ferromagnet.
\end{lemma}

\begin{proof} Indeed, we have
\begin{equation*}
d\nu_{\delta,g}^{\Pi} = Z_{\Pi}^{-1}
\exp\left(-\frac{1}2\sum_{i,\,{i'}\in\Pi}
N_{i,\,i'}^{(\Pi,\,l)}t_it_{i'}\right)\prod_{j\in\Pi}\,d\nu_j\,,
\end{equation*}
where
\begin{equation*}
d\nu_j = \exp\left\{\lambda\,g(x_j)\,t_j\right\} \exp\left\{
{-\wick{Q(t_j)}(g(x_j))}\right\}\,dt_j\,.
\end{equation*}
Note that the numbers $-N_{i,\,i'}^{(\Pi,\,l)}$ are nonnegative
for $i\ne i'$, since the non-diagonal elements of the matrix
$N^{(\Pi,\,l)}$ are non-positive which was proven above. All the
rest is provided by nonnegativeness of $\lambda$ and $g$, and by
evenness of $Q$.
\end{proof}

Next we state a version of Griffiths' theorem (see Ref. 13,
Theorem VIII.3) for even Ising ferromagnets.

\begin{theorem}[Griffiths] Let the measure $d\mu$ correspond to an
even Ising ferromagnet, and let
$\xi^i=\xi_1^{i_1}\ldots\xi_r^{i_r}$. Then
\begin{description}
\item[(a)] (Griffiths' first inequality)\qquad
$\int\xi^i\,d\mu(\xi)\ge 0$;%
\smallskip\nobreak
\item[(b)] (Griffiths' second inequality)\qquad
$\int\xi^{i_1+i_2}\,d\mu\ge \left(\int\xi^{i_1}\,d\mu\right)
\left(\int\xi^{i_2}\,d\mu\right)$.
\end{description}
\end{theorem}

It follows from the last theorem that if $\{\,h_1,\ldots,h_r\}$ is
a collection of nonnegative functions from $\DlPi$, then the
Schwinger function $\SchdgPi(h_1,\ldots,h_r)$ is nonnegative.
Indeed, since $h_i\ge 0$, it is sufficient to show, that
$\SchdgPi(e_{i_1},\ldots,e_{i_r})\ge 0$, but that follows from the
Griffiths' first inequality and lemma~4.1.

\subsection{Monotonous increase}
\noindent \looseness1 Here we show the monotonous increase of the
Schwinger functions. As noted above, it is sufficient to consider
the lattice approximations for that functions, where the
``lattice'' consists of balls $O_i$ of arbitrarily small radius
$\delta = q^l$.

\begin{theorem}
Let the polynomial $P$ be of form $P(X)=Q(X)-\lambda X$, where
$Q(X)$ is an even polynomial, and $\lambda$ is a nonnegative
constant. Let then $\Pi,\Pi'$ be the unions of $\eta$ and $\eta'$
balls of the radius $q^l$, respectively, such that
$\Pi\subset\Pi'$. Finally, let $g$ be a nonnegative function from
$\DPi$, and $\{\,h_1,\ldots,h_r\}$ be a collection of nonnegative
functions from $\DPi$. Then
\begin{equation*}
%\label{eqn4.13}%
\SchdgPi\,(h_1,\ldots,h_r)\le
\SchdgPiC\,(h_1,\ldots,h_r).
\end{equation*}
\end{theorem}
\begin{proof} Since $h_i\ge 0$, it is sufficient to establish the
inequality $$\SchdgPi\,(e_{k_1}^{(l)},\ldots,e_{k_r}^{(l)})\le
\SchdgPiC\,(e_{k_1}^{(l)},\ldots,e_{k_r}^{(l)}),$$ 
$e_{k_i}^{(l)}\in \DPi$. By the formula
(\ref{eqn4.8}) we can write
\begin{multline*} \SchdgPi\,(e_{k_1}^{(l)},\ldots,e_{k_r}^{(l)})  \\ =
Z_{\Pi}^{-1}\int\limits_{{\mathbb R}^\eta} t_{k_1}\cdots x_{k_r}
\exp\left(-\frac{1}2\sum_{i,\,{i'}\in\Pi}
N_{i,\,i'}^{(\Pi,\,l)}t_it_{i'}\right)\prod_{j\in\Pi}e^{-\wick{P(t_j)}(g)}\,d^\eta
t  \\ =  Z^{-1}\int\limits_{{\mathbb R}^{\eta'}} t_{k_1}\cdots
t_{k_r}  \exp\left(-\frac{1}2\sum_{i,\,{i'}\in\Pi}
N_{i,\,i'}^{(\Pi,\,l)}t_it_{i'}
 -\frac{1}2\sum_{i,\,{i'}\in\Pi'\setminus\Pi}
N_{i,\,i'}^{(\Pi',\,l)}t_it_{i'}\right) \times \\ \times
\prod_{j\in\Pi'}e^{-\wick{P(t_j)}(g)}\,d^{\eta'} t,
\end{multline*}
where
\begin{multline*}
Z=\int\limits_{{\mathbb R}^{\eta'}}\exp\left(-\frac{1}2\sum_{i,\,{i'}\in\Pi}
N_{i,\,i'}^{(\Pi,\,l)}t_it_{i'}
 -\frac{1}2\sum_{i,\,{i'}\in\Pi'\setminus\Pi}
N_{i,\,i'}^{(\Pi',\,l)}t_it_{i'}\right) \\ \times
\prod_{j\in\Pi'}e^{-\wick{P(t_j)}(g)}\,d^{\eta'} t,
\end{multline*}
so that the integral with respect to the variables
$\{t_j\}_{j\in\Pi'\setminus\Pi}$ reduces because of the
normalization multiplier $Z^{-1}$. 

Using the equality
(\ref{eqn4.5}) we can rewrite $\SchdgPi\,(e_{k_1}^{(l)},\ldots,e_{k_r}^{(l)})$ as
\begin{multline*} Z^{-1}\int\limits_{{\mathbb R}^{\eta'}} t_{k_1}\cdots
t_{k_r} \exp\left(-\frac{1}2\sum_{i,\,{i'}\in\Pi}
N_{i,\,i'}^{(\Pi',\,l)}t_it_{i'}
-\frac{1}2\sum_{i,\,{i'}\in\Pi'\setminus\Pi}
N_{i,\,i'}^{(\Pi',\,l)}t_it_{i'}\right) \times \\ \times
\prod_{j\in\Pi'}e^{-\wick{P(t_j)}(g)}\,d^{\eta'} t.
\end{multline*} It is
important that the renormalization of the Wick polynomial $P$ is
taken with respect to the free measure $\mu_0$ and does not depend
on the region $\Pi$ or $\Pi'$.

The function $\SchdgPiC\,(e_{k_1}^{(l)},\ldots,e_{k_r}^{(l)})$ has the same
form as $\SchdgPi\,(e_{k_1}^{(l)},\ldots,e_{k_r}^{(l)})$, except that the sum
$\left(\sum_{i,\,{i'}\in\Pi} N_{i,\,i'}^{(\Pi',\,l)}t_it_{i'}
+\sum_{i,\,{i'}\in\Pi'\setminus\Pi}
N_{i,\,i'}^{(\Pi',\,l)}t_it_{i'}\right)$ is added by the term
$\sum\limits_{i\in\Pi,\ {i'}\in\Pi'\setminus\Pi}
N_{i,\,i'}^{(\Pi',\,l)}t_it_{i'}$, and the normalization multiplier is
changed appropriately. We put
\begin{multline*}
F(\tau)\\ = Z_{\,\tau}^{-1} \int\limits_{{\mathbb R}^{\eta'}}
t_{k_1}\cdots t_{k_r} \exp\left(-\frac{1}2\sum_{i,\,{i'}\in\Pi}
N_{i,\,i'}^{(\Pi',\,l)}t_it_{i'}
-\frac{1}2\sum_{i,\,{i'}\in\Pi'\setminus\Pi}
N_{i,\,i'}^{(\Pi',\,l)}t_it_{i'}\right)  \\ \times
\prod_{j\in\Pi'}e^{-\wick{P(t_j)}(g)} \exp\left(-\
\tau /2\sum_{i\in\Pi,\ {i'}\in\Pi'\setminus\Pi}
N_{i,\,i'}^{(\Pi',\,l)}t_it_{i'}\right)\,d^{\eta'} t,
\end{multline*}
where $Z_{\,\tau}$ is the corresponding normalization multiplier,
and $0\le\tau\le 1$. Then, we have
$F(0)=\SchdgPi\,(e_{k_1}^{(l)},\ldots,e_{k_r}^{(l)})$, and
$F(1)=\SchdgPiC\,(e_{k_1}^{(l)},\ldots,e_{k_r}^{(l)})$. Let us show that
$F(1)\ge F(0)$. We have
\begin{multline}
\label{eqn4.14} \frac{dF}{d\tau}\\
=\frac{1}2\sum\limits_{i\in\Pi,
{i'}\in\Pi'\setminus\Pi}
\left(-N_{i,\,i'}^{(\Pi',\,l)}\left\{\,\langle\, t_{k_1}\cdots
t_{k_r}\ t_i\,t_{i'}\,\rangle_\tau - \langle\, t_{k_1}\cdots
t_{k_r}\,\rangle_\tau\, \langle\,
t_it_{i'}\,\rangle_\tau\right\}\, \right),
\end{multline}
where $\langle\ \cdot\ \rangle_\tau$ denotes the mean with respect
to the measure
\begin{multline*}
d\mu_{\,\tau} = Z_{\,\tau}^{-1} \exp\left(-\frac{1}2\sum_{i,\,{i'}\in\Pi}
N_{i,\,i'}^{(\Pi',\,l)}t_it_{i'} +\frac{1}2(1-\tau)\sum_{i\in\Pi,\
{i'}\in\Pi'\setminus\Pi} N_{i,\,i'}^{(\Pi',\,l)}t_it_{i'}\right)
\\ \times \prod_{j\in\Pi'}e^{-\wick{P(t_j)}(g)}\,d^{\eta'}
t.
\end{multline*}
It is easy to see that  the measure $d\mu_{\,\tau}$ corresponds to
an even Ising ferromagnet. So the derivative (\ref{eqn4.14})
is nonnegative, since $-N_{i,\,i'}^{(\Pi',\,l)}\ge 0$ for
$i\in\Pi,\ {i'}\in\Pi'\setminus\Pi$ by the property of
non-diagonal elements of the matrix $N^{(\Pi',\,l)}$ proven near
the beginning of subsection~4.2, and the expression enclosed in
braces in the right-hand part of~(\ref{eqn4.14}) is
nonnegative by the Griffiths' second inequality.
\bigskip
\end{proof}

\medskip
\section*{Acknowledgment} \noindent This research was
supported in part by CRDF under Grant UM1--2421--KV--02.

\end{document}